# Optimizing Web Sites for Customer Retention*

Michael Hahsler

Department of Information Systems and Operations

Vienna University of Economics and Business Administration

**Abstract**

*With customer relationship management (CRM) companies move away from a mainly product-centered view to a customer-centered view. Resulting from this change, the effective management of how to keep contact with customers throughout different channels is one of the key success factors in today's business world. Company Web sites have evolved in many industries into an extremely important channel through which customers can be attracted and retained. To analyze and optimize this channel, accurate models of how customers browse through the Web site and what information within the site they repeatedly view are crucial. Typically, data mining techniques are used for this purpose. However, there already exist numerous models developed in marketing research for traditional channels which could also prove valuable to understanding this new channel. In this paper we propose the application of an extension of the Logarithmic Series Distribution (LSD) model repeat-usage of Web-based information and thus to analyze and optimize a Web Site's capability to support one goal of CRM, to retain customers. As an example, we use the university's blended learning web portal with over a thousand learning resources to demonstrate how the model can be used to evaluate and improve the Web site's effectiveness.*

## 1. Introduction

Today, the World Wide Web is a very important channel for attracting new customers and for retaining customers. For example, a company Web site can be used for conveying service and product information, for advertising and for selling products. Optimizing the Web appearance and integrating it into an overall customer relationship management (CRM) strategy is a key success factor in today's business world. Therefore, it is crucial to gain an insight into how customers use the Web site.

Analyzing large volume Web usage data is normally done by data mining specialists using their own techniques. However, there exist numerous models developed in marketing research for traditional channels which could prove valuable to understand this new channel. It is important to examine how well existing models perform on such data and, if necessary, to adapt the models accordingly. A further challenge is to make the application of the models efficient enough so they can be used together with other data mining techniques for large amounts of data (e.g., by simplification or by applying computationally more efficient estimation techniques from machine learning).

In marketing research, a well-known group of models for repeat-buying behavior are the family of NBD-type models. Such a model was first introduced for consumer goods by Chatfield et al. [1] and later extended and applied by various authors. For example, Lee et al. [2] used the NBD model to describe the visit frequencies of users for a Web site and gave an empirical illustration using the visit frequencies extracted from the transaction log of the *Yahoo!* Web portal. Moe and Fader [3] modeled the visit frequencies of visitors for the online retailers *Amazon* and *CDNOW* with an NBD-type model they extended to incorporate a stochastic process of changes in visit frequencies over time. On their data sets from an online panel collected directly by the visitors' Web browsers, the new model provides a slightly better fit than a stationary NBD model. Notable is that their model can split the visitors into a group with decreasing visit rates and a group with increasing rates, where in their data set frequent visitors with an increasing visit frequency show an up to 40% higher conversion rate than the other visitors. These two applications of the model are analogous to brick-and-mortar store visiting behavior which is known to be covered well by NBD-type models (see e.g. [4]).

In this paper we focus on the application of a NBD-type model for repeat usage at the level of individual Web pages or collections of Web pages that belong together and serve a common purpose (e.g., product descriptions, support information) rather than on the visits of whole Web sites. With this approach it is possible to analyze and compare different parts of a Web site, differences between products or product classes in Web-based catalogues, or different ways of organizing and presenting information and services. Also, analyses of the impact of other CRM activities, promotions, and changes in the customer behavior over time are possible.

To simplify model estimation and thus make it more suitable for larger data volumes, we suggest using the Logarithmic Series Distribution (LSD) model, a simplification of the zero-truncated NBD model. We also extend the repeat usage model to include one-time users which seem to play an important role in Web usage data.

The paper is organized as follows: In section 2 we discuss differences between Web-based information and consumer products to analyze how models originally developed for consumer products can be applied to Web-based information. In section 3 the LSD model is presented and we extend the model to account for one-time users. In section 4 we present the empirical results for fitting the model to a data set from an educational portal. In section 5 we describe how the model can be used by managers to analyze and optimize parts of their Web site.

## 2. Web-based Information *vs.* Consumer Products

The NBD model and the repeat-buying theory were developed for consumer products and not for information. In this section we compare consumer goods with so called 'information goods' and motivate why NBD-type models can also be applied to such information goods. Shapiro and Varian [4] suggest defining *information* very broadly by stating that everything that can be digitized is information. Therefore examples of *information goods* are books, music, movies, stock quotes, and so on. This definition clearly also comprises online product or support information, downloads, etc. In the online world an information good can be a collection of Web pages which belong together (information is spread over different Web pages for better access or includes additional media in separate files). Furthermore, online information goods can be dynamic, they can be personalized or consist of search results for a user's query.

Similar to consumer products, information goods can be offered, sold and consumed. For the model used in this paper an important characteristics of consumer goods is that they are consumed in regular intervals (e.g., a family buys some milk almost every week). Many information goods create similar usage patterns, for example a person might check her cell phone provider's Web site almost every week to see what new ring tones or logos are available.

**Table 1: Differences between using consumer panels and Web usage data for NBD-type models.**

|  | **Consumer Panel** | **Web Usage Data** |
|---|---|---|
| **Items** | consumer products | information goods (Web sites, collections of Web pages, dynamic content, multimedia content, etc.) |
| **Unit of Analysis** | market baskets | browser sessions |
| **Analyzed Behavior** | purchase incidences | following links to Web sites or documents, downloads, etc. |
| **Ignored Behavior** | quantity bought, package size | repeat usage per session, number of pages browsed within an information good, time spent on a page |
| **Identity of Customers** | known | varies between anonymous user sessions and complete personalization |
| **Purchase History** | known | unknown or partially known |
| **Non-buyers** | known | unknown |
| **Incorrect Recording** | omission, over-reporting | dynamic IP addresses, caching mechanisms, proxy servers, Web robots |

The data originally used for the NBD model stem from consumer panels. Although, there exist online panels which use modified Web browsers to record the user behavior (e.g. comScore Media Metrix), this approach is very cost intensive and it probably only provides enough information for the analysis of the most popular Web sites. The most common and inexpensive way to collect consumer data is analyzing the transaction data logged by the company's Web server. Most of the differences between panels and transaction logs are just the names used to refer to the same concepts (see Table 1), but there are also some fundamental differences. For example, in panels the whole purchase history of each customer is known, for anonymous use of information on Web sites this is usually not the case. Even if the Web site uses personalization, often the whole usage history of the user cannot be reconstructed since users tend to use a site anonymously as long as possible and only reveal their identity when it is absolutely necessary. Another important difference is the errors known to occur during data recording. Even though data recording at the level of an application server (content server, e-commerce system, etc.) is becoming more common, still online data recording is far from perfect. Known problems include user and session identification with dynamic IP addresses, caching mechanisms and proxy servers as well as the presence of Web robots (e.g., see [6] and [7]). The NBD-type models are used on consumer panel data which are also known to be imperfect and contain omission and over-reporting of purchases [8] which produce similar distortions like online data recording. The reason why the model still works reasonably well lies in its inherent robustness which is a very important property for analyzing Web usage data.

## 3. The LSD/OTB Model

In this section we develop a simple and robust NBD-based model for the usage frequencies of information goods which uses the LSD approximation of the NBD extended by users who only access a specific information once and then never again. Although we model usage, in the spirit of repeat-buying theory, we refer to such users as one-time "buyers" (OTB). We start with an introduction of the NBD-type model and then derive the LSD/OTB model. We also review the assumptions of the NBD model to argue why the model is appropriateness for the data analyzed in this paper and under which circumstances the model can or can not be applied.

Chatfield, Ehrenberg and Goodhardt introduced in [1] a simple but very successful model of stationary purchasing behavior, the so-called negative binomial distribution (NBD) model, which Ehrenberg [8] used to describe repeat-buying behavior for consumer products. The model uses the following simplifications:
1. Only one brand (product) is analyzed at a time. Therefore, the relations between different brands (products) are irrelevant and brand choice and product heterogeneity are not modeled. This simplification carries over unchanged to online data where only one information product is analyzed at a time.
2. The aggregated purchasing behavior of all customers is stable (stationary) during the analyzed period of time, meaning there is no trend present. This has been found to hold approximately for frequently bought consumer goods, however, it certainly does not hold for the introduction of new products. This assumption can be problematic since online information goods can be short-lived. However, since the production cost of high quality information is high, these information goods will be present long enough to reach a near stationary market after an introduction phase. Furthermore, the stationary model can be used as a reference to detect trends in a non-stationary situation.
3. The model uses the concept of purchasing occasions and purchase incidences. A purchase incidence occurs, if a consumer purchases one or more units of a product in a specific trip to a store (a purchasing occasion). The number of items bought and the package size are ignored. The frequency of purchases proved to be more useful for modeling than the quantity bought or the price paid. Applied to information goods this means that repeat visits per session, the number of pages browsed within an information good and the time spent on these pages, is ignored.

For the NBD model we consider one specific product for a single time period $t$. Set $C$ represents the customers. The purchases of the product for each individual customer $c_i \in C$ are modeled by Poisson processes with parameters $\mu_{ci}$:

$$P_{c_i}(R=r) = \frac{\mu_{c_i}^{-r} e^{-\mu_{c_i}}}{r!} \ for \ r = 0, 1, 2, ...$$

Customer heterogeneity is accounted for by different parameters $\mu_{ci}$. The distribution of the long-run $\mu_{ci}$ over all customers is assumed to follow a truncated Γ-distribution. The Γ-distribution has 3 parameters and can fit various shapes. Following this assumption, the distribution of the number of purchases per customer is a special case of a compound Poisson distribution, namely the negative binomial distribution (NBD). See [8] for the proof.

The NBD has 2 parameters, the mean *m* and the exponent *k*. The following equation gives the probability of observing a customer purchasing the product *r* times in period *t*.

$$P_{NBD}(R=r) = \left(1 + \frac{m}{k}\right)^{-k} \frac{\Gamma(k+r)}{\Gamma(r+1)\Gamma(k)} \left(\frac{m}{m+k}\right)^r \; for \; r = 0, 1, 2, ...$$

The mean *m* is the average number of purchases per customer. $P_{NBD}(R=0)$ is the proportion of active customers who just do not buy the analyzed product in the observed period of time. The proportion of customers *b* who buy is therefore *1- $P_{NBD}(R=0)$* and the average number of purchases these customers is *ω =m/b*.

The NBD can be fitted to empirical data using the method of moments, maximum likelihood estimation and some other estimation methods [9], but in general parameter estimation is cumbersome. Especially, if the proportion of customers who do not buy the item $P_{NBD}(R=0)$ is not observable, which is often the case for Web usage data. However, the zero-truncated NBD tends (for *k<0.2*) toward the logarithmic series distribution (LSD) with only one parameter, *q* (see [10]). The LSD gives a very good approximation to the NBD if the proportion of customers who buy the item in the population is less then 20% (*b<0.2*) [8]. This is almost always the case for Web usage data with a large number of items. For the LSD, the probability of observing customers buying *r*-times in period *t* is given by:

$$P_{LSD}(R=r) = \frac{-q^r}{r \ln(1-q)} \; for \; r = 1, 2, ...$$

Where the average number of purchases per customer *ω* is a function of *q*:

$$\omega = \frac{-q}{(1-q)ln(1-q)}$$

The LSD is easy to fit to empirical data. For the maximum likelihood estimation the mean ω is estimated by the sample mean and then the equation above is used to compute the estimate for the parameter *q*. The parameter *q* cannot be calculated directly from *ω*, however, standard numeric approximation can be applied efficiently.

In some situations, real-world data sets contain a proportion of observations at *r=1* which is much higher than expected. Since the model only describes repeat-usage behavior, the discrepancy at *r=1* can be explained by users who only use a certain information product once and then never again. Fader and Hardie [11] incorporated such one-time buyers (OTB) into the NBD model and develop the NBD/OTB model to describe purchases in a tea store. In the same way we expand the LSD model to model one-time users. The aggregate distribution of purchases of the LSD/OTB model is:

$$P_{LSD/OTB}(R=r) = \pi \delta_{r,1} + (1-\pi) P_{LSD}(R=r) \; for \; r = 1, 2, ...$$

$\pi$ is the proportion of one-time buyers and $\delta_{r,1}$ is the Kronecker delta with $\delta_{r,1} =1$ if $r=1$ and $0$ otherwise. The mean and variance are given by:

$$E(R) = \pi + (1-\pi)\omega$$

$$Var(R) = (1-\pi)\frac{\omega}{1-q} + \pi(1-\pi)(1-2\omega - \frac{1-\pi}{\pi}w^2)$$

If the proportion $\pi$ is 0 then the LSD/OTB model collapses into the simple LSD model. The parameters of the LSD/OTB model can be estimated using the method of maximum likelihood. The log-likelihood function is given by:

$$LL(q,\pi) = \sum_{r=1}^{max(r)} f_r ln(P_{LSD/OTB}(R=r|q,\pi))$$

$f_r$ represents the observed count of users with $r$ repeat-buys. To estimate the parameters, standard numerical optimization is used to maximize the log-likelihood function. Alternatively, the real number of repeat-users with $r=1$ could be seen as unknown and it could be estimated iteratively together with the distribution parameter $q$ using the *Expectation Maximization* (EM) algorithm [12].

## 4. Fitting the Model

To provide evidence that the LSD/OTB model can describe usage of information from a Web site, we analyze one year (2001) of transaction data from the application level log of an educational portal, the Virtual University information system at the Vienna University of Economics and Business Administration (see http://vu.wu-wien.ac.at). The portal provides access to online information (lecture notes, research material, enrollment information, etc.) for students and researchers. An item in this setting is defined as a collection of Web pages (including also other media like word processor files) that present a logical unit of information on a specific topic, e.g., *Course material on International Marketing*. The items are stored on different Web servers in the university's Intranet as well as on the Internet. The portal provides an interface to find items and records when a user follows the hyperlink to an item. The system is comparable to a Web-based catalogue or a small, very specialized search engine.

Since the portal does not require user identification (which is also common for commercial Web sites), we have to use heuristics (filter Web robots, analyze IP addresses and use permanent cookies stored in the users' Web browsers) to identify different users [6]. To compute sessions, we used a timeout of 20 minutes (a common timeout for this purpose, see [13]). This is a very coarse approach because of dynamic IP addresses and proxy servers. However, we will see that this data still provides useful insights for understanding aggregated Web usage behavior.

**Table 2: Results of fitting the LSD and the LSD/OTB models***

| model | items | no q | no $\chi^2$ test ($\alpha$=0.05) | no significant differences | significant differences | % of fitting models |
|---|---|---|---|---|---|---|
| **LSD** | 2226 | 713 | 786 | 475 | 252 | **65.34%** |
| *LSD/OTB* | *2226* | *834* | *838* | *516* | *38* | ***93.14%*** |

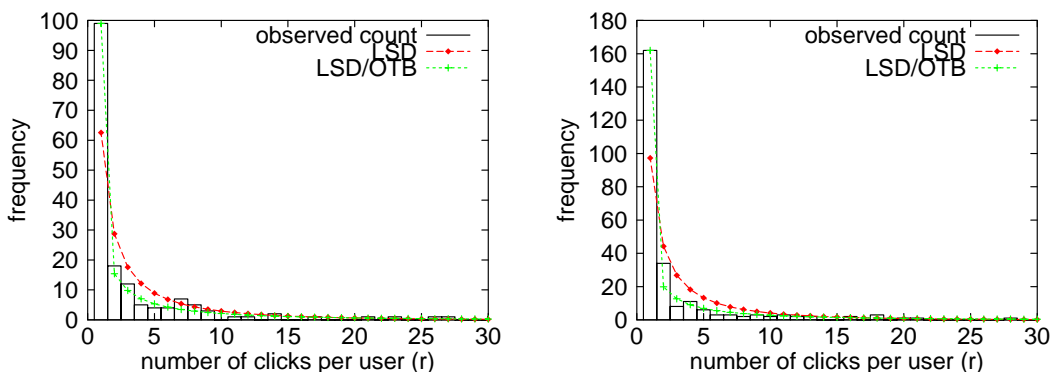

**Figure 1: Two examples (*C++ Standard Library* to the left and *Freshmeat* to the right). For better visibility of the differences both plots are truncated at r=30.**

In the data set we observed usage of 9671 different items from 15340 different IP addresses. The data contained 56278 sessions. We selected again items with more than 10 observations. This reduced the items to 2226. For these items we estimated the parameters of the LSD and LSD/OTB model and used the $\chi^2$-goodness-of-fit test to analyze the fit of the model for our data. The results of the analysis are shown in Table 2. For the items with enough observations and repeat-usage to estimate the parameter *q* and to conduct the $\chi^2$ test, the observations of 65.34% of the items do not differ significantly from the LSD model. The introduction of one-time users improves the results. For the LSD/OTB model more than 90% of the items do not differ significantly from the model.

In Figure 1 we inspect 2 items in more detail. The left plot is for an item which contains information about the *C++ Standard Library* for programming in the language C++. The LSD/OTB model (p=0.954, $\pi$=0.390) fits the data well with $\chi^2$=1.922 at 5 degrees of freedom (no significant differences at $\alpha$=0.05). Following the model, 39% of the clicks are from one-time users who only needed the information once or found the pages not informative so they never returned. The simple LSD model is significantly different from the observed data since it can not fit the high number of observations at *r=1*.

The plot to the right in Figure 1 is for the item *Freshmeat,* a repository of free software. The LSD/OTB model (q=0.956, $\pi$=0.471) describes the observed data better than the LSD model, however, the $\chi^2$-value is 15.134 at 6 degrees of freedom which means that there are significant differences between the model and the data. The reason for the differences are that in the observed data there are several users who use the site more than 30 times (this is not visible in Figure 1 since the plot is truncated at r=30). These observations distort the estimates for the parameters which leads to a underestimation at r=2 and 4. An explanation for the observed high repeat-users could be recording errors due to the user identification heuristic.

---

* Numbers are corrected in this version of the paper.

## 5. Using the LSD/OTB Model for Analyzing Web Sites

The parameters of the model directly provide important diagnostics for the usage of different items. A proportion of one-time users (parameter $\pi$) which is considerably higher for some items than for the rest in a Web Site indicates that the items do not provide the consumer with the expected information or are not fulfilling their expectations, thus do not contribute to the aim of retaining customers. An analysis of these items or the navigation structure for these items can provide information to improve the Web site or the item and thus help to convert one-time users to repeat-users.

**Table 3: Parameters of the LSD/OTB model over time for 3 selected items**

|  | 1st Quarter | | | 2nd Quarter | | | 3rd Quarter | | | 4th Quarter | | |
|---|---|---|---|---|---|---|---|---|---|---|---|---|
|  | user | q | $\pi$ | user | q | $\pi$ | user | q | $\pi$ | user | q | $\pi$ |
| Enrolment System | 67 | 0.72 | 0.79 | 20 | 0.87 | 0.91 | 46 | 0.70 | 0.79 | 27 | 0.96 | 0.95 |
| C++ Class | 111 | 0.81 | 0.36 | 93 | 0.84 | 0.56 | 36 | 0.80 | 0.95 | 26 | 0.99 | 0.75 |
| Online Dictionary | 241 | 0.92 | 0.51 | 213 | 0.90 | 0.41 | 161 | 0.87 | 0.30 | 173 | 0.83 | 0.31 |

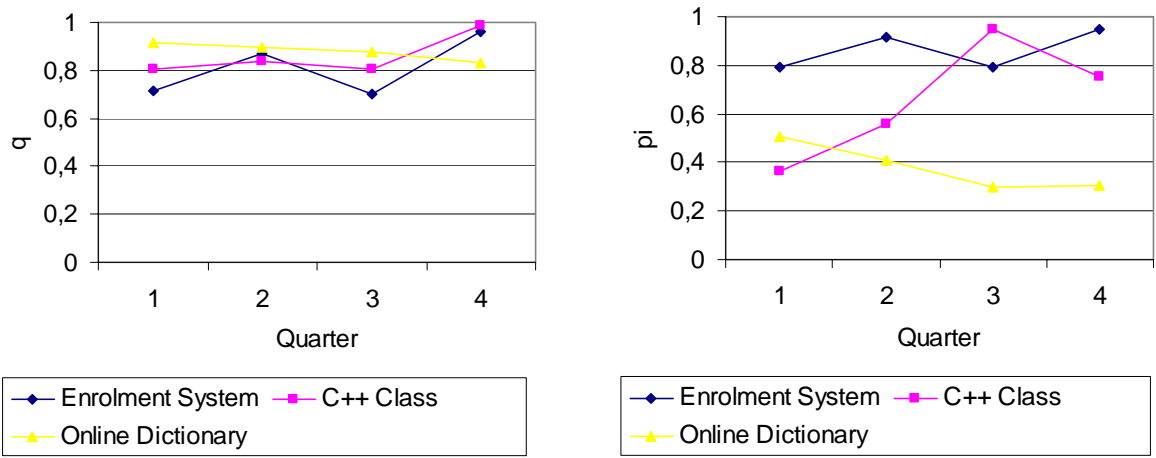

**Figure 2: Parameters of the LSD/OTB model over time**

The parameter *q* can be interpreted as the proportion of usage by repeat-users. By considering more than one time-period with the LSD model, this repeat-usage behavior can be analyzed. We divided the data set containing data for 1 year into 4 quarters and estimated the LSD/OTB model for each quarter individually. In Table 3 we present the number of different users who used the items and the estimated model parameters for three selected items. The items are the university wide online enrolment system, an introductory class on the programming language C++ and a online English-German dictionary. The changes of the parameters over the 4 quarters are shown in the plots in Figure 2.

For the strictly stationary LSD model, the values of *q* should not change over time unless there are changes in the underlying user behavior. The usage behavior for the online dictionary exhibits approximately stationary behavior. However, parameter *q* declines steadily which means the item is losing repeat-users. For the remaining two items we found greater changes of the parameter *q* over time. The changes of both items are similar and can be explained by seasonality in the data set. The 3$^{rd}$ quarter (July to September) are holidays at the university where only few students are on campus and with the 4$^{th}$ quarter (October) the new study year starts. Therefore, we observe for the 2 items a lower *q* in the 3$^{rd}$ quarter and a very high *q* in the 4$^{th}$ quarter.

Changes of the parameter *π*, the proportion of one-time users, over time also indicates changes in usage behavior. In the data for the selected 3 items (see Figure 2) no clear trend can be identified. For the enrolment system the proportion seams to stay almost constant at a very high rate. This might indicate that navigating to the system through the portal page is cumbersome compared to the prominently placed link on the university's home page. For the online dictionary *π* it is steadily degreasing. For the C++ class we observed the biggest changes which can be explained by the behavior of its main user group, students who can take the class. The 1$^{st}$ quarter is dominated by students studying for the final exam in February, hence the low one-time only user rate. The class is not offered in the 2$^{nd}$ and the 3$^{rd}$ quarter resulting in a higher *π* with the peak in the 3$^{rd}$ quarter where many students check out if they want to take the class in the next term starting with the 4$^{th}$ quarter.

If stationary usage behavior (a constant number of users and a constant parameter *q*) can be assumed approximately the model can be used for predicting repeat-usage (analogous to repeat-buying in [8]). The theory considers two consecutive periods of the same length where by definition the parameter *q* stays constant. Individual users change their behavior but the aggregated numbers stay the same. In this situation users who use an item in both periods are called *repeat-users* (repeat-buyers). *New users* (new buyers) use an item in period two but not in period one. *Lost users* (lost buyers) used an item in the first period but do not use it in the second period. From the parameter *q* of the LSD model several characteristics of repeat-usage behavior can be calculated. First, the proportion of repeat-users $b_R$ in the total number of observed users *b* can be calculated by:

$$\frac{b_R}{b} = q + \frac{ln(1+q)}{ln(1-q)}$$

Note that this proportion is more than just the number of users who used an item more than once during the observation period. It also contains a proportion of the users who used the item only once in the observation period but are predicted by the model to use the item again in the following period.

The proportion of times an item is used by repeat-users $m_R$ in each time period equals *q*:

$$\frac{m_R}{m} = q$$

The average usage per repeat-user $\omega_R$ is given by:

$$\omega_R = \frac{q^2}{ln(1-q)ln(1-q^2)}$$

And finally, the average number of times an item is used by lost or new users (they are equal in the stationary state) can be calculated by:

$$\omega_L = \omega_N = \frac{q}{ln(1+q)}$$

Table 4: Repeat-usage characteristics for 3 selected items

|  | $q$ | $\omega_R$ | $b_R/b$ | $m_R/m$ | $\omega_L=\omega_N$ |
|---|---|---|---|---|---|
| Enrolment System | 0.717 | 2.519 | 0.572 | 0.717 | 1.327 |
| C++ Class | 0.807 | 3.205 | 0.640 | 0.807 | 1.364 |
| Online Dictionary | 0.916 | 5.456 | 0.737 | 0.916 | 1.409 |

We calculated the corresponding values using $q$ estimated from the 1st quarter and present the results in Table 4. For the C++ class the average repeat-user uses the class Web pages 3.205 times in the 1st quarter. 64.0% of the observed users (excluding the one-time users) are repeat-users who account for 80.7% of the usage of the class pages. The new and lost users on average use the pages 1.364 times, a number close to 1. In the same way the numbers for the other items can be interpreted where the online dictionary has very loyal repeat-users (73.7% of its users) accounting for over 90% of the observed usage.

Even if stationary or near-stationary behavior cannot be assumed (e.g., for the C++ class), the above characteristic are still meaningful to analyze the composition of the user base from usage data. The values are more useful than simple statistics normally generated by Web analyzer tools. Standard analyzer tools report summary statistics as the total number of page views, visits and hits in a time period. These numbers can give an idea about the development of the usage over several periods but the structure of repeat-usage is not visible. Clearly, retained customers, i.e., customers who repeatedly visit some Web pages are more valuable for commercial Web sites than customers who never come back after the first visit. With the LST/OTB model observed usage from a single time period can be divided into a part resulting from one-time users and another part resulting from repeat users. This provides a very important insight into how effective parts of the Web site are in converting visitors into loyal repeat-visitors.

## 6. Conclusion

In this paper we concentrated on analyzing usage-behavior of information provided by a commercial Web sites in order to work towards one of the main CRM goals, customer retention. Instead of applying standard data mining technology, we applied a NBD-type model from marketing research to model usage frequencies of collections of Web pages (Web-based information). In order to be able to deal with larger volumes of data, we applied the LSD model, the most simple and easy to estimate version of the NBD-type models. To account for one-time users (OTB), we expanded the model and developed the LSD/OTB model. Using an example data set, we show that the LSD/OTB model gives a very good fit for the items (for over 90% of the items).

The parameters of the extended model are directly usable for the manager of a Web site. For example, big differences between the proportion of one-time users within a Web site give a hint to problems with the item quality or the navigational structure (e.g., an item is reachable by a link with a misleading name). Changes of the users' repeat-usage behavior over time can be also automatically analyzed by detecting changes in the parameters in successive periods. These insights into the composition of usage-behavior are way beyond the capabilities of standard Web analyzer tools.

Especially, the estimation of the proportions of repeat-users and one-time users directly shows the ability of different parts of the Web site to retaining customers and thus indicates where improvements are required. This is a valuable input for the CRM processes.

## 7. References

[1] C. Chatfield, A. S. C. Ehrenberg, and G .J. Goodhardt, "Progress on simplified model of stationary purchasing behaviour," *Journal of the Royal Statistical Society, Series A*, 129(3):317-367, 1966.

[2] S. Lee, F. Zufryden, and X. Dreze, "Modeling consumer visit frequency on the internet," In *34th Annual Hawaii International Conference on System Sciences (HICSS-34)-Volume* 7, 2001.

[3] W.W. Moe, and P. S. Fader, "Capturing Evolving Visit Behavior in Clickstream Data," Marketing Science Conference, 2003.

[4] G.A. Grisbie, and Frisbie, Jr, "Ehrenberg's Negative Binomial Model Applied to Grocery Store Trips, " *Journal of Marketing Research,* Vol. 17, August 1988, pp.385-390.

[5] C. Shapiro, and H. R. Varian, *Information Rules*, Harvard Business School Press, Boston, MA, 1999.

[6] R. Cooley, B. Mobasher, and J. Srivastava, "Data preparation for mining world wide web browsing patterns," *Journal of Knowledge and Information System*s, vol. 1, no. 1, 1999.

[7] P.-N. Tan and V. Kumar, "Discovery of web robot sessions based on their navigational patterns," *Data Mining and Knowledge Discover*y, vol. 6, pp. 9–35, 2002.

[8] A. S. C. Ehrenberg, *Repeat-Buying: Facts, Theory and Applicatio*n. London: Charles Griffin & Company Ltd., 1988.

[9] R.C. Tripathi, Negative binomial distribution. In Samuel Kotz and Norman L. Johnson, editors, *Encyclopedia of Statistical Science*, volume 6, pages 169-177. John Wiley & Sons, New York, 1985.

[10] G.P. Patil, Logarithmic series distribution. In Samuel Kotz and Norman L. Johnson, editors, *Encyclopedia of Statistical Science*, volume 5, pages 111-114. John Wiley & Sons, New York, 1985.

[11] P. S. Fader and B.G.S. Hardie, "A note on an integrated model of customer buying behavior," *European Journal of Operational Research*, 139(3):682-687, 2002.

[12] A. P. Dempster, N. M. Laird, and D. B. Rubin. Maximum likelihood from incomplete data via the EM algorithm. *Journal of the Royal Statistical Society,* Series B (Methodological), 39:1–38, 1977.

[13] L. D. Catledge and J. E. Pitkow, "Characterizing browsing strategies in the World-Wide Web," *Computer Networks and ISDN System*s, vol. 27, no. 6, pp. 1065–1073, 1995.